\documentclass{article}

\usepackage{PRIMEarxiv}

\usepackage[utf8]{inputenc} 
\usepackage[T1]{fontenc}    
\usepackage{hyperref}       
\usepackage{url}            
\usepackage{booktabs}       
\usepackage{amsfonts}       
\usepackage{nicefrac}       
\usepackage{microtype}      
\usepackage{lipsum}
\usepackage{fancyhdr}       
\usepackage{graphicx}       
\graphicspath{{media/}}     
\usepackage[version=3]{mhchem}
\title{Laminar flow synthesis of submicron \ce{CaCO3} particles in 3D printed microfluidic chips

}

\author{
	Reznik I.A. \\
	International Research and Education Centre\\ 
	for Physics of Nanostructures\\
	ITMO University\\
	197101 Saint Petersburg, Russia\\
	\\
	Faculty of Electrical Engineering and Computing\\
	University of Campinas\\
	13083-970 Campinas, Brazil\\
	Corresponding author: \\
	\texttt{ivanreznik1993@mail.ru} \\
	\And
	Kolesova E.P. \\
	International Research and Education Centre\\ 
	for Physics of Nanostructures\\
	ITMO University\\
	197101 Saint Petersburg, Russia\\
	\\
	Center of translation medicine\\
	Sirius University\\
	354349 Sochi, Russia\\
	\And
	Pestereva A.S.\\
	International Laboratory\\
	Hybrid Nanostructures for Biomedicine\\
	ITMO University\\
	199034 Saint Petersburg, Russia \\
	\And
	Baranov K.N.\\
	International Laboratory\\
	Hybrid Nanostructures for Biomedicine\\
	ITMO University\\
	199034 Saint Petersburg, Russia \\
	\And
	Osin Y.N.\\
	Laboratory for Scientific Restoration of Precious Metals\\
	The State Hermitage museum\\
	191186 Saint Petersburg, Russia\\
	\And
	Bogdanov K.V.\\
	International Research and Education Centre\\ 
	for Physics of Nanostructures\\
	ITMO University\\
	197101 Saint Petersburg, Russia\\
	\And
	Swart J.W.\\
	Faculty of Electrical Engineering and Computing\\
	University of Campinas\\
	13083-970 Campinas, Brazil\\
	\And
	Moshkalev S.A.\\
	Centre for Semiconductor Components and Nanotechnology\\
	University of Campinas\\
	13083-870 Campinas, Brazil\\
	\And
	Orlova A.O.\\
	International Laboratory\\
	Hybrid Nanostructures for Biomedicine\\
	ITMO University\\
	199034 Saint Petersburg, Russia \\
}

\begin{document}
\maketitle

\begin{abstract}
Microfluidic technology provides a solution to the challenge of continuous \ce{CaCO3} particle synthesis. In this study, we utilized a 3D-printed microfluidic chip to synthesize \ce{CaCO3} micro and nanoparticles in vaterite form. Our primary focus was on investigating a continuous one-phase synthesis method tailored for the crystallization of these particles. By employing a combination of confocal and scanning electron microscopy, along with Raman spectroscopy, we were able to thoroughly evaluate the synthesis efficiency. This evaluation included aspects such as particle size distribution, morphology, and polymorph composition. The results unveiled the existence of two distinct synthesis regimes within the 3D-printed microfluidic chips, featuring a channel cross-section of 2 mm\(^{2}\). In the first regime, characterized by turbulence, particles with an average diameter of around 2 \(\mu\)m were produced, displaying a broad size distribution. Conversely, the second regime, marked by laminar flow, led to the synthesis of submicron particles (approximately 800-900 nm in diameter) and even nanosized particles (70-80 nm). This research significantly contributes valuable insights to both the understanding and optimization of microfluidic synthesis processes, particularly in achieving controlled production of submicron and nanoscale particles.
\end{abstract}


\section{Introduction}
Drug delivery for cancer treatment represents a pivotal area of research aimed at enhancing therapeutic efficacy while minimizing systemic side effects \cite{hafeez2021challenges, yao2020nanoparticle}. The Enhanced Permeability and Retention (EPR) effect, intrinsic to the unique characteristics of tumor vasculature induced, including by hypoxia, allows for preferential accumulation of nanoparticles within the tumor microenvironment \cite{sun2022tumor, zi2022strategies}. Utilizing nanoparticle-based drug delivery systems capitalizes on the EPR effect to achieve targeted drug delivery, optimizing drug concentrations at the tumor site and mitigating off-target effects \cite{sharifi2022updated}. Despite significant advancements, only a small number of nanoparticles in medicine reach the stage of clinical trials and receive approval \cite{parodi2022anticancer}. Micro- and nanoparticle platforms based on calcium carbonate (\ce{CaCO3}) show promise for advancing drug delivery in photodynamic therapy \cite{maleki2015calcium}.
To minimize off-targeting effect, controlled-release systems from a matrix responsive to external conditions like temperature, enzyme activity or pH emerge as a viable approach \cite{egorova2023smart, maleki2015calcium}. The capability of degrading porous \ce{CaCO3} nanoparticles layer by layer as the pH level decreases provides control over the release rate of molecules from subsequent layers of the porous matrix, as well as the oxygen enrichment time for further excitation in cancer cells/tumors \cite{maleki2015calcium, zhong2016degradation}. Additionally, the particle porosity can also control the rate of \ce{CaCO3} degradation and payload release. Most traditionally used techniques produce micron-sized particles (size) while reducing the size of nanocarriers can significantly increase the drug delivery efficiency by improving biodistribution, circulation time and cellular internalization \cite{sukhorukov2004porous, antipov2003carbonate, lu2009size, panyam2003biodegradable}.
Microfluidic technologies are well-established and widely used for chemical synthesis of nanomaterials in various fields, such as chemistry, materials science, biomedicine, and environmental sciences \cite{he2018intensified, tofighi2019microfluidically, shi2019recent, kung2019microfluidic, tichit2019synthesis}. Microfluidic systems offer significantly higher surface-to-volume ratios compared to conventional flask-based systems, resulting in better control over heat and mass transfer rates. Literature analysis shows that microfluidic synthesis outperforms flask-based synthesis in terms of controllability, monodispersity, reproducibility, reagent efficiency, and impurity minimization, thanks to better control over nanocrystal growth processes. 
Microfluidic synthesis using single-phase laminar flow combined with flow rate control can effectively enhance the synthesis efficiency and size distribution of nanoparticles. The process involves initiating efficient reactions and using appropriate control mechanisms to produce desired nanomaterials (e.g., controlling mixing efficiency, pH, temperature, etc.). However, designing, optimizing, and integrating functional components for synthesis processes present significant challenges.
In this paper we developed a procedure to effectively slow down \ce{CaCO3} particle growth, offering better control of particle synthesis. Key to our study is the design and 3D printing of microfluidic chips with 2 mm2 and 0.5 mm2 channel cross-sections. These chips facilitated the synthesis of both micron and submicron-sized \ce{CaCO3} particles, showcasing the efficiency of this cutting-edge technology.
Through CFD simulations, we conducted an in-depth analysis of synthesis dynamics, revealing two distinct regimes. The first, turbulent regime produced micron-sized particles, while the second, laminar regime resulted in submicron particles. 

\section{Materials and Methods}
\label{sec:headings}

\subsection{Materials Used}

The materials utilized in this study were as follows: Calcium Chloride (Sigma Aldrich, St. Louis, MO, USA), Sodium carbonate (Vekton, Saint-Petersburg, Russia), Deionized water, Ethylene Glycole (Ekos 1, Moscow, Russia), Ethanol, Photopolymer resin (Anicubic, China). The water for experiments was purified using a Milli-pore system. The rest of the chemicals were used without further purification.

\subsection{Instruments and Techniques for Investigating the Physical, Chemical, and Optical Properties of Samples}

Size distribution of synthesized \ce{CaCO3} particles were analyzed through Scanning Electron Microscopy (SEM). SEM measurements were conducted using a scanning electron microscope SU 7000 HITACHI equipped with EDS detectors BRUKER (XFlash 6160, XFlash Flat QUA5060F). The specimen was deposited on a silicon substrate with a subsequent application of a 7nm conductive layer of Cr. The application of the conductive layer was performed using the cathodic sputtering method in the vacuum setup EMACE600 - LEICA.
For the analysis of the morphology and sizes of nanoparticles, both LD and UD detectors were employed (LD - lower SE detector, UD - upper intra-lens SE detector). The accelerating voltage was set at 5 keV. Elemental quantitative analysis and the construction of element distribution maps of nanoparticles were carried out at an accelerating voltage of 15 keV using EDS detectors BRUKER XFlash 6160. The probing depth at 15 keV is 1 micron. 
Morphology of synthesized particles was analyzed with Raman spectroscopy. The raman spectra were obtained on an InVia micro-Raman spectrometer (Renishaw, UK). The measurements were carried out under excitation by an Ar+ laser with a wavelength of 514 nm. The InVia spectrometer operates under backscattering conditions and has a Leica microscope with an x50 objective with Na=0.75 and a multichannel detector (CCD camera) cooled to -70°C. The area of irradiation with exciting light is ~2 \(\mu\)mm.

\subsection{Microfluidic Chip Fabrication}
For the synthesis of \ce{CaCO3} microspheres, microfluidic  chips were prepared according to the procedure described in the work already published by our lab \cite{baranov2023optical}. The microfluidic chip consists of [N×M] matrix replicated basic cells. Each basic cell was characterized by three parameters: Channel Width (Ch-W) ranging from 0.5 to 2 mm, Wall Width (W-W) of 0.8 mm, and Channel Height (Ch-H) of 1 mm. Connector cells were used to link the columns of basic cells. Based on this logic, a script was developed for the parametric generation of microfluidic chips using the OpenScad software package. 
To create physical replicas of the chips using additive manufacturing methods, the chip design was subtracted from a solid block, thus forming the microfluidic channel cavities. The direct printing of microfluidic chips was carried out using the Anicubic Photon Mono (Anicubic, China) photopolymer 3D printer, utilizing generic photopolymer resin (Anicubic, China). 

Printing parameters for photopolymer resin were as follows: the thickness of one layer was 50 \(\mu\)m, and each layer was exposed to UV light (405 nm) for 2 seconds. After printing, a significant amount of unpolymerized resin remained within the chip channels and on its surface. To remove this residue, the chip was placed in an ultrasonic bath filled with isopropyl alcohol for 5-10 minutes. Subsequently, the internal chip channels were rinsed with pure isopropyl alcohol using a syringe pump connected to the chip's outlet.

The connection of the syringe pumps to the microfluidic chip was achieved using Teflon tubes with internal and external diameters of 1 mm and 1.5 mm, respectively, and steel adapters with internal and external diameters of 0.9 mm and 1.1 mm, respectively. The chip connection process was performed in two stages. In the first stage, a steel adapter was inserted at one end into a Teflon tube and at the other end into an inlet or outlet hole of the microfluidic chip. In the second stage, to ensure chip sealing, all connections were coated with multiple layers of photopolymer resin, followed by UV light exposure at a wavelength of 405 nm for 10-30 seconds to polymerize the joints.

\subsection{Computational fluid dynamics simulations}
The 2D chip designs were created using the OpenScad software package and subsequently imported into COMSOL Multiphysics 5.2 as a foundation for computational fluid dynamics simulations. The simulations utilized the “Laminar flow (spf)” and “Transport of Diluted Species (tds)” interfaces. The solver was configured to handle time-dependent equations. To optimize computational efficiency in the CFD process, the mesh size for the stenosis section was set to 10 \(\mu\)m.

The obtained results were employed to assess the variation in fluid mixing dynamics between two distinct chip designs, each having channel widths of 0.5 and 2 mm, respectively. The model initially contained water, and reagent solutions were introduced from the inlet at the microchannel's edge with a flow rate of 1 \(\mu\)L/second. A zero-gauge pressure condition was established at the outlet.

For the simulation of \ce{CaCl2} and \ce{Na2CO3} reagent solutions, both inlets were configured to introduce water, defining two separate aqueous phases with viscosities, densities, and diffusion constants corresponding to 0.33M solutions. The resulting phase composition was graphically represented. Table 1 provides a summary of all constants utilized in the CFD calculations.

\begin{table}[h!]
 \caption{Parameters of Precursors and Microfluidic Setup for the CFD calculations of \ce{CaCO3} reagents mixing}
  \centering
  \begin{tabular}{lll}
    \toprule
    Parameter     & \ce{CaCl2}     & \ce{Na2CO3} \\
    \midrule
    Viscosity, mPa·s &  \multicolumn{2}{c}{1.0016}  \\
    Density, g/mL     & 1.0279 & 1.0207     \\
    Diffusion coefficient, m\(^2\)/s \(\times\)10\(^{-9}\)    & 1.134      & 1.11 \\
    Concentration, mole/m\(^3\) & \multicolumn{2}{c}{330}  \\
    Temperature, \(^\circ C\) & \multicolumn{2}{c}{25}  \\
    Volumetric flow rate*, ml/s & \multicolumn{2}{c}{0.001}  \\
    \bottomrule
  \end{tabular}
  \label{tab:table}
\end{table}

\subsection{Synthesis of \ce{CaCO3} Particles}
In all experiments, aqueous solutions of \ce{CaCl2} and \ce{Na2CO3} with concentrations of 0.33 mol/L each were used as precursors. The components were mixed in a 1:1 ratio either in an Eppendorf tube using a magnetic stirrer for 100 seconds or in a microfluidic chip for the same amount of time. Total volume if reagent mixture used for each synthesis type was set to 200 \(\mu\)L. Mixing time duration in microfluidic chip was controlled by using microfluidic chip with inner volume of 100 \(\mu\)L and speed of reagent flow set to 0.001 mL/s. After mixing, the reagent mixture and resulting calcium carbonate particles were diluted 2, 5, and 10 times using distilled water. These solutions, along with the undiluted mixture, were then cleansed twice from unreacted reagents by centrifugation at a speed of 5000 rpm for one minute. Subsequently, the precipitate was redissolved in 100 \(\mu\)L of water and drop casted onto the surface of a glass slide for further analysis.

\section{Results and discussion}
\subsection{Passive inhibition of \ce{CaCO3} Synthesis }
The synthesis of calcium carbonate particles is widely used for its simplicity and rapidity, requiring only the mixture of reactants for initiation and continuation of the reaction process \cite{boyjoo2014synthesis, sovova2021calcium}. These minimal requirements make the transition from bulk flask synthesis to continuous microfluidic synthesis relatively straightforward. Initially, we conducted a comparative analysis of the size distribution and morphology of microparticles synthesized using both flask and microfluidic methods. To ensure comparability, synthesis conditions were closely matched: Both approaches employed the same reagent concentration of 0.33 M and a synthesis duration of 100 seconds. In the case of microfluidic synthesis, the reaction time is governed by the fluid velocity required to traverse the inner volume of the microfluidic chip within the specified timeframe (exact parameters and principal schematic of microfluidic chip used in this experiment are given in supporting information). Figure 1 displays microscope images, size distribution profiles, and Raman spectra of the resultant particles.

\begin{figure}[h!]

\centering

\begin{minipage}{.4\textwidth}
  \centering

    (\textbf{a})

  \includegraphics[width=1.1\linewidth]{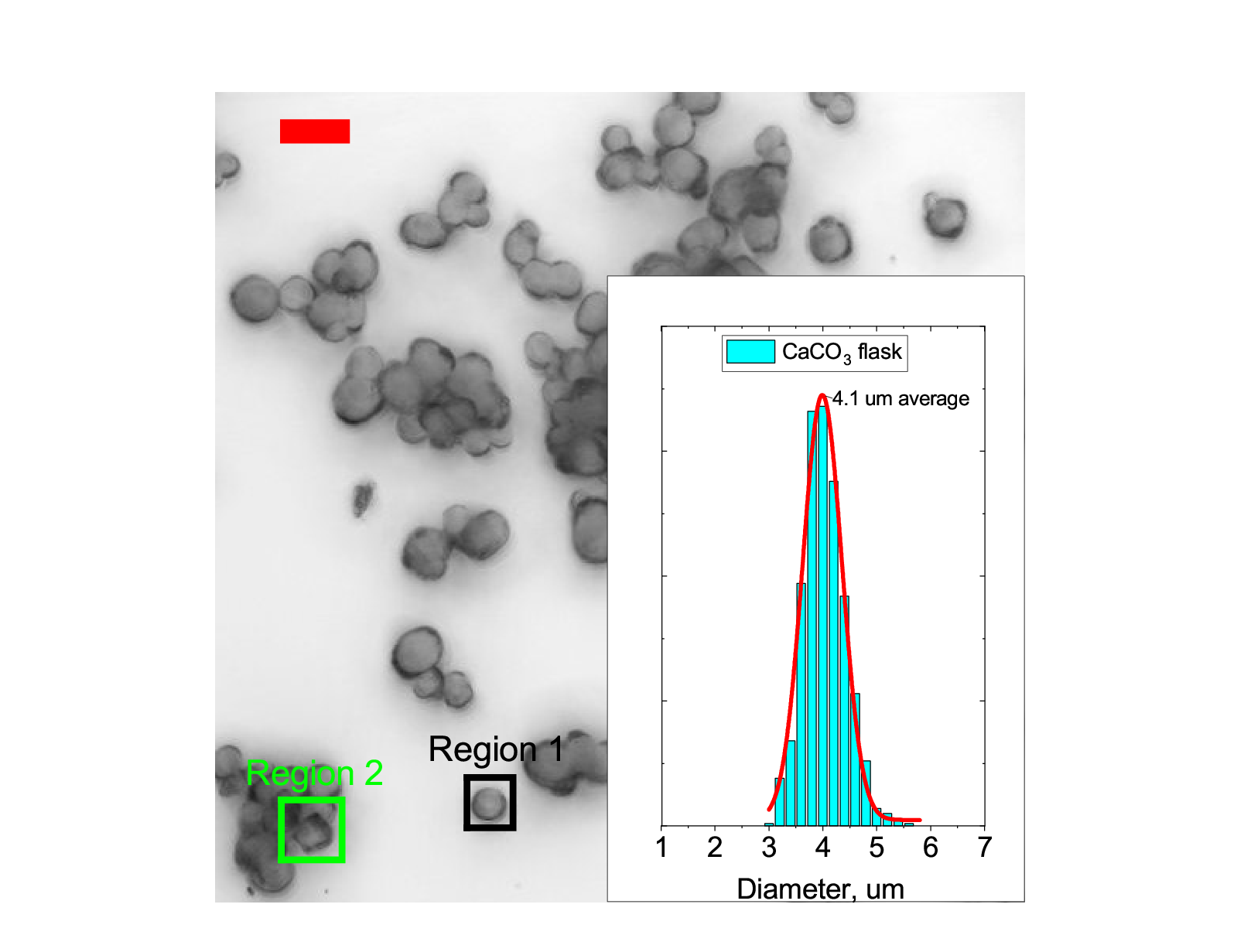}

  \label{fig:1_a}
\end{minipage}%
\begin{minipage}{.4\textwidth}
  \centering

    (\textbf{b})

  \includegraphics[width=1.15\linewidth]{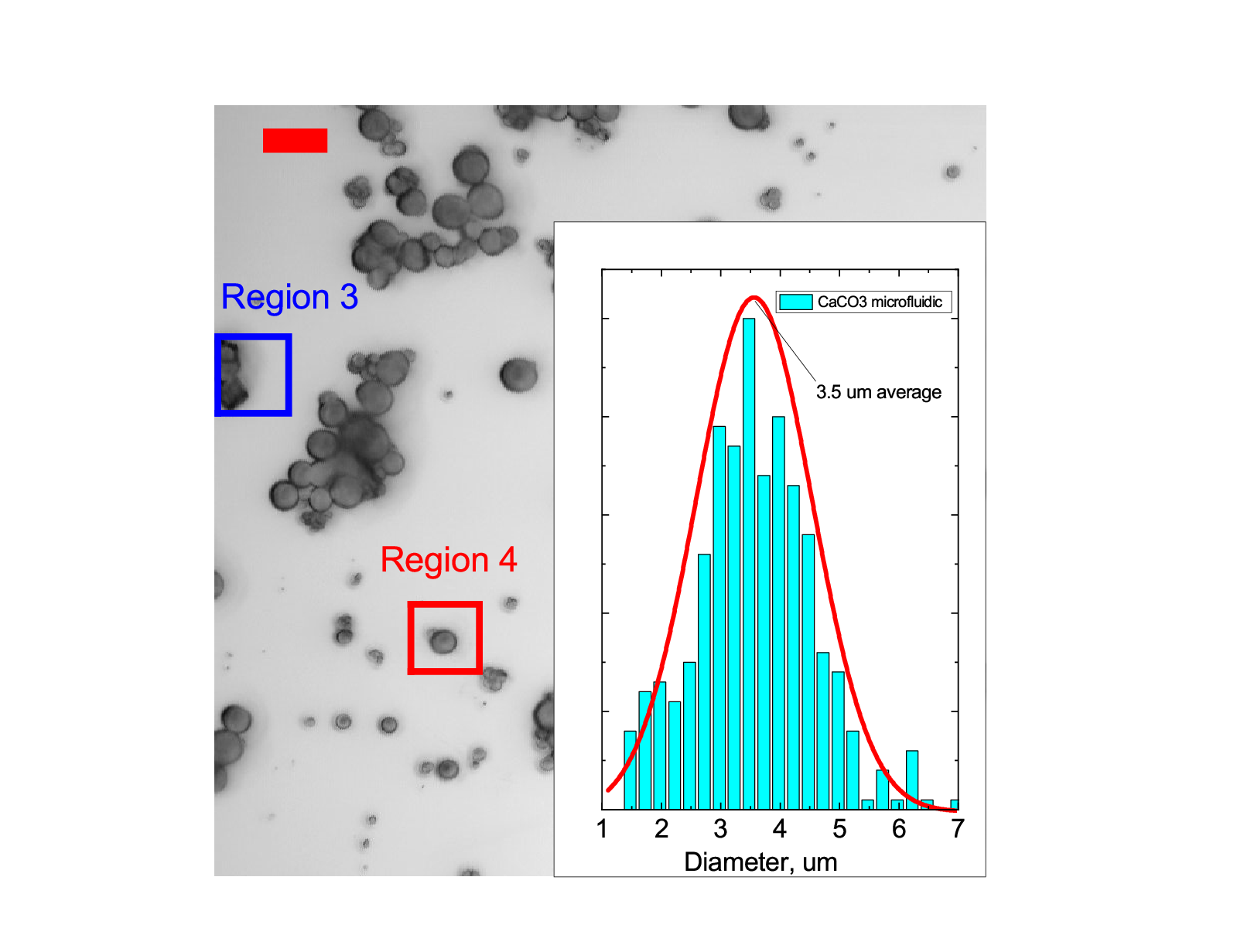}
  
  \label{fig:1_b}
\end{minipage}

\vspace{10pt}
\centering

    (\textbf{c})

  \includegraphics[width=0.7\linewidth]{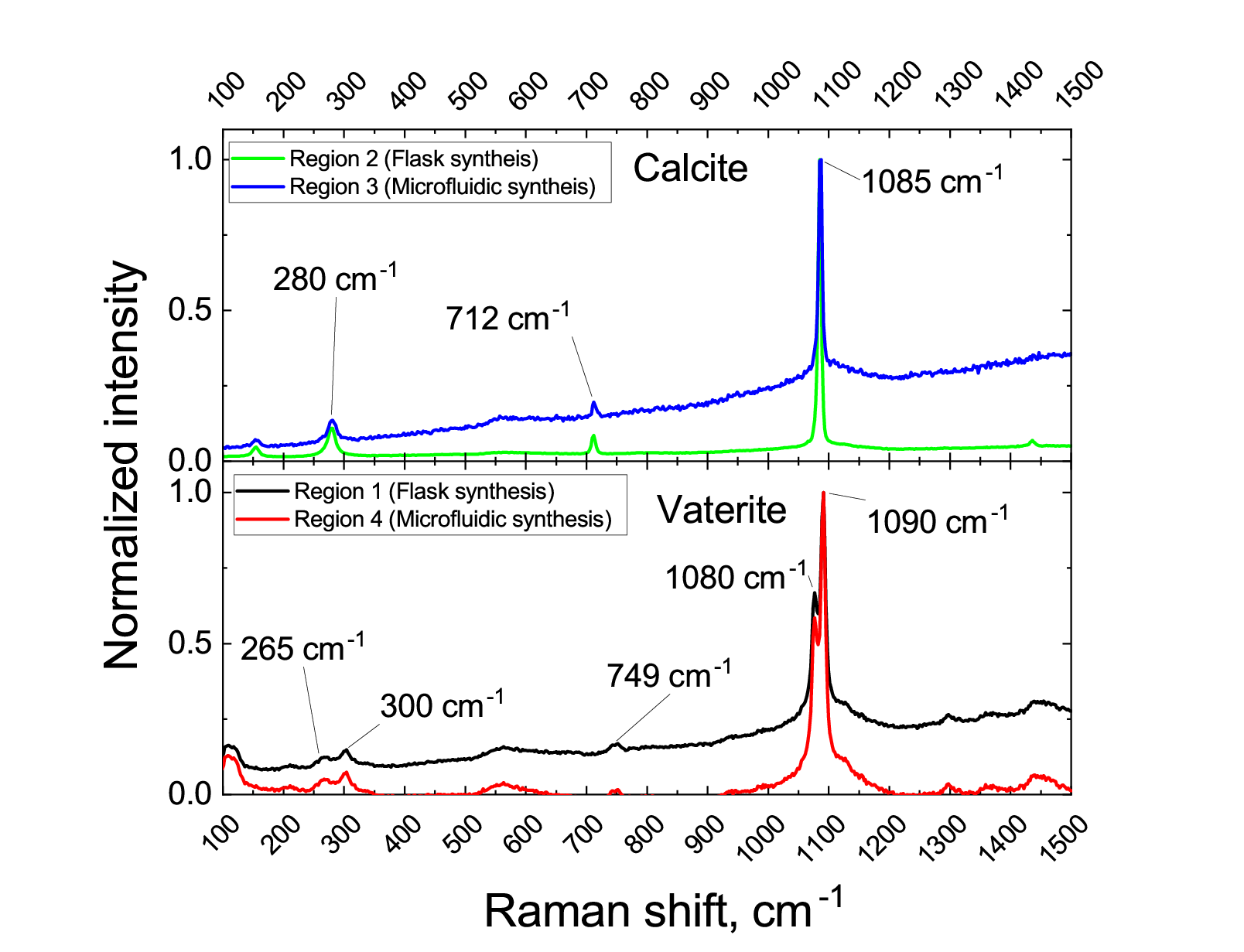}

  \label{fig:1_c}

\caption{(\textbf{a},\textbf{b}) $-$ microscope images of \ce{CaCO3} particles synthesized in a flask and microfluidic chip, respectively. Bar length on images is 5 micrometers. Inset - size distribution of respective synthesis batch \ce{CaCO3} particles. (\textbf{c}) $-$ Raman spectra of \ce{CaCO3} particles taken from sites designated on microscope images.
}
\label{fig:1}
\end{figure}

From Figure 1, it is evident that both synthesis methods (flask and microfluidic chip) yield similar outcomes. In both cases, particles formed with diameters around 4 micrometers. Although particles produced through the microfluidic route exhibit a slightly smaller average diameter (3.5 \(\mu\)m compared to 4.1 \(\mu\)m for flask synthesis), their size distribution is notably broader (2.3 \(\mu\)m for microfluidic versus 1.6 \(\mu\)m for flask synthesis). This variation might arise due to non-uniform mixing of reagents during their transport along the microfluidic channels. 
As can be seen from Figures 1a and 1b, \ce{CaCO3} particles represented predominantly by spherical form.  A minor fraction of particles is observed in the shape of a cube (Figure 1a and 1b, region 2 and 3). Analysis of the Raman spectra of the samples synthesized in the bulk and microfluidic chip, as depicted in Figure 1c, revealed characteristic single peaks at 1080 and 1090 cm\(^{-1}\) corresponding to Vaterite form of \ce{CaCO3}. As well as the 1085 cm\(^{-1}\)  peak that corresponds to Calcite form of \ce{CaCO3} \cite{donnelly2017synthesis}.  Singlet peaks at 1080 and 1090 cm\(^{-1}\) arise from aginternal mode due to the \(\nu_1\)-symmetric stretching mode of the carbonate ion. The \(\nu_4\) in-plane bending mode of carbonate can be found at 712 cm\(^{-1}\) (calcite) and 749 cm\(^{-1}\) (vaterite). Finally, both samples exhibit peaks below 300 cm\(^{-1}\), which correspond to the translational and rotational lattice modes. 
It is noteworthy that while microfluidically synthesized \ce{CaCO3} microparticles exhibit a relatively wide size distribution, they also exhibit the potential for further diameter reduction that lies in a precise control of reagents reaction time. However, due to the absence of specific external reaction initiators, there is no alternative but to reduce the concentration of reactant molecules in the particle growth region of the mixture to decrease the reaction rate and/or completely halt the reaction \cite{chaussemier2015state}. Therefore, at the next step of the study, we evaluated the particle growth dynamics depending on the concentration of reactant molecules in the precursor mixture. For this purpose, immediately after mixing the reactants in the flask and before precipitating the resulting particles using centrifugation, the precursor mixture was diluted 5 and 10 times. The concentrations of the synthesis reactants and the particle precipitation protocols from the reaction mixture are provided in Section 2.5 Materials and Methods. Figure 2 displays confocal microscopy images of the resulting particles and their size distribution, deposited on glass slides, at various degrees of precursor mixture dilution before precipitation.

\begin{figure}[h!]

\centering

\begin{minipage}{.5\textwidth}
  \centering

 \hspace{-1.2cm}   (\textbf{a})

 \hspace{-1.2cm}   \includegraphics[width=1.0\linewidth]{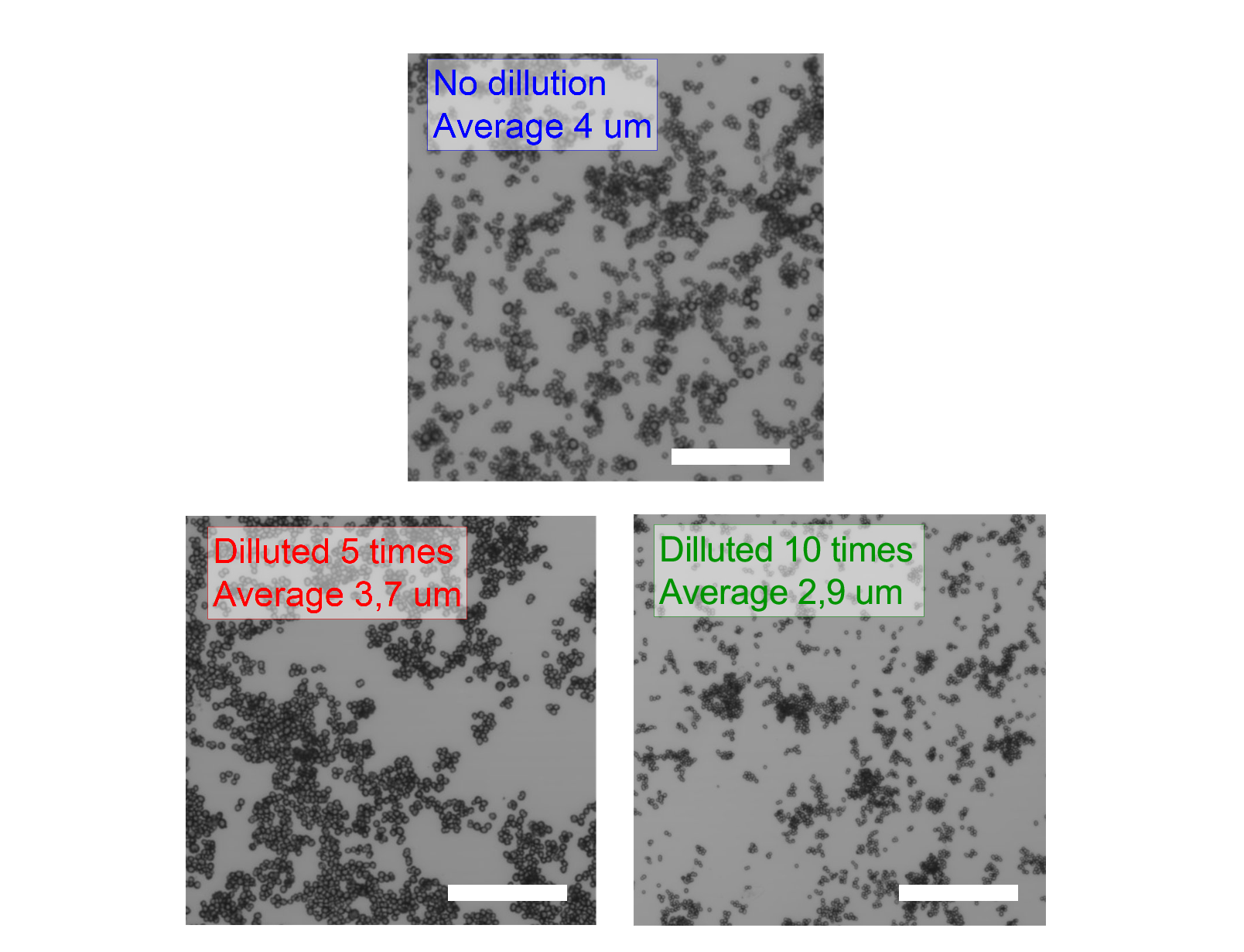}

  \label{fig:2a}
\end{minipage}%
\begin{minipage}{.5\textwidth}
  \centering

    (\textbf{b})

  \includegraphics[width=1.0\linewidth]{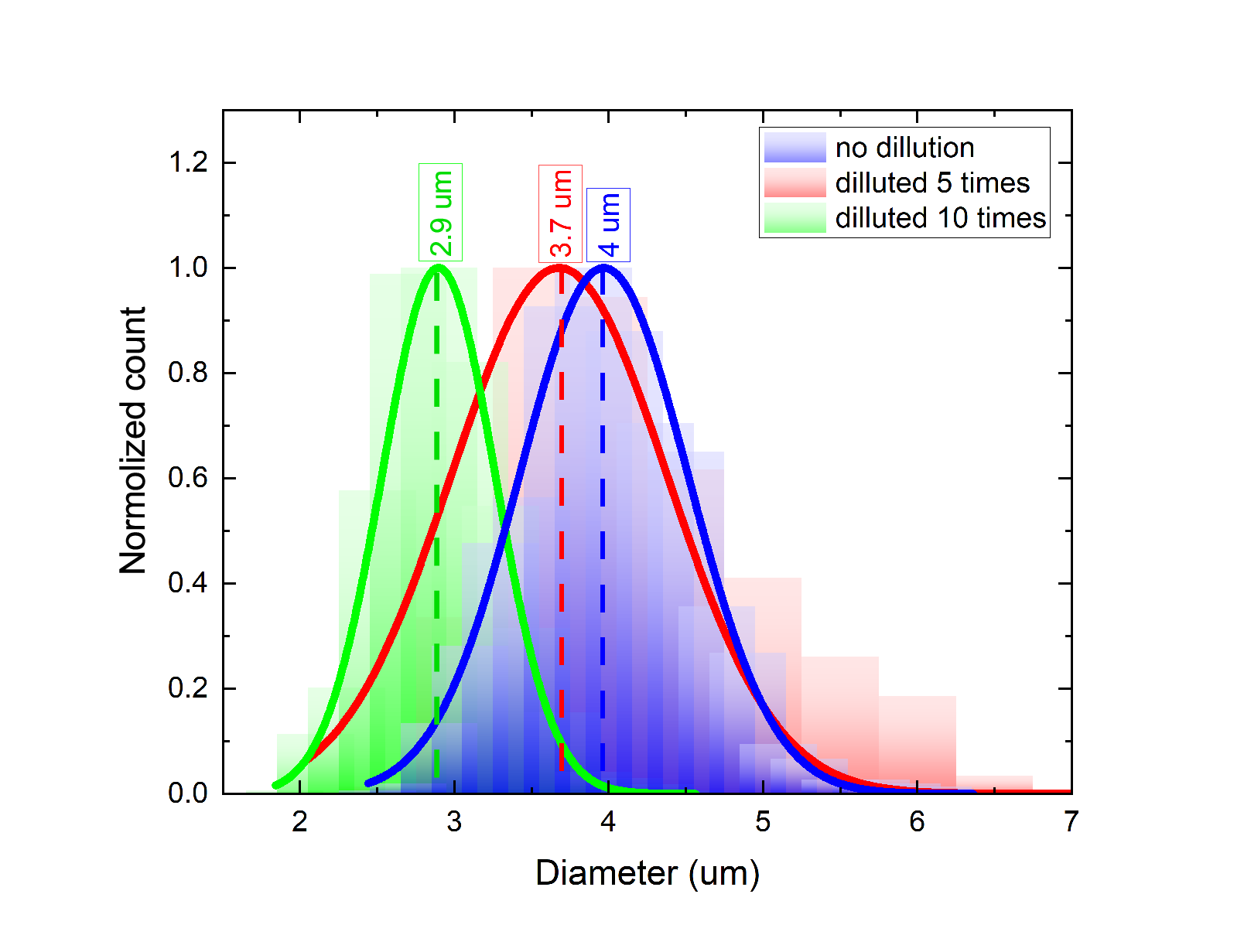}
  
  \label{fig:2b}
\end{minipage}
\caption{Confocal images (\textbf{a}) and normalized size distribution plots (\textbf{b}) of \ce{CaCO3} particles, diluted by a factor of 5 (red curve), 10 (green curve), as well as undiluted (blue curve) prior to precipitation from the precursor mixture. To determine the peak position of the particle size distribution, the size distribution curves were fitted with a Gaussian function. A 20-micrometer scale bar is provided in the confocal images.
}
\label{fig:2}
\end{figure}

Analysis of the particle size distributions of calcium carbonate, as depicted on the Figure 3b, led to the conclusion that precipitation of particles from the reaction mixture without dilution allows the formation of particles with a diameter of \(\sim \)4 \(\mu\)m and a width of size distribution (FWHM) of 1.5 \(\mu\)m. Dilution of the reactant mixture by a factor of 5 and 10 led to a reduction in particle diameter to 3.7 \(\mu\)m (FWHM - 1.5 \(\mu\)m) and 2.9 \(\mu\)m (FWHM - 0.85 \(\mu\)m), respectively. The increase in the size of calcium carbonate particles by a factor of two when the reaction mixture was diluted by two-fold before precipitation is likely attributed to additional reagent mixing along with relatively moderate concentration dilution. Conversely, further reduction in precursor concentrations in the reaction mixture results in an additional reduction in particle size compared to the original undiluted mixture. This particle growth dynamic is likely caused by a sharp change in pH level from pH 9.6 to pH 7 in the reaction mixture. Such transitioning from alkaline to neutral conditions leads to particle size reduction \cite{oral2018influence}. Based on the obtained data, it can be inferred that dilution of the reaction mixture by ten or more times is an effective strategy not only for halting the reaction of calcium carbonate particle growth but also for reducing their final size. 
Additionally, we investigated the kinetics of calcium carbonate particle growth as a function of the degree of dilution of the reaction mixture prior to precipitation. Figure 3 presents the results of the analysis of the average size of the obtained calcium carbonate microspheres, depending on the degree of precursor mixture dilution and time after mixing. 

\begin{figure}[h!]
\centering
  \includegraphics[width=0.9\linewidth]{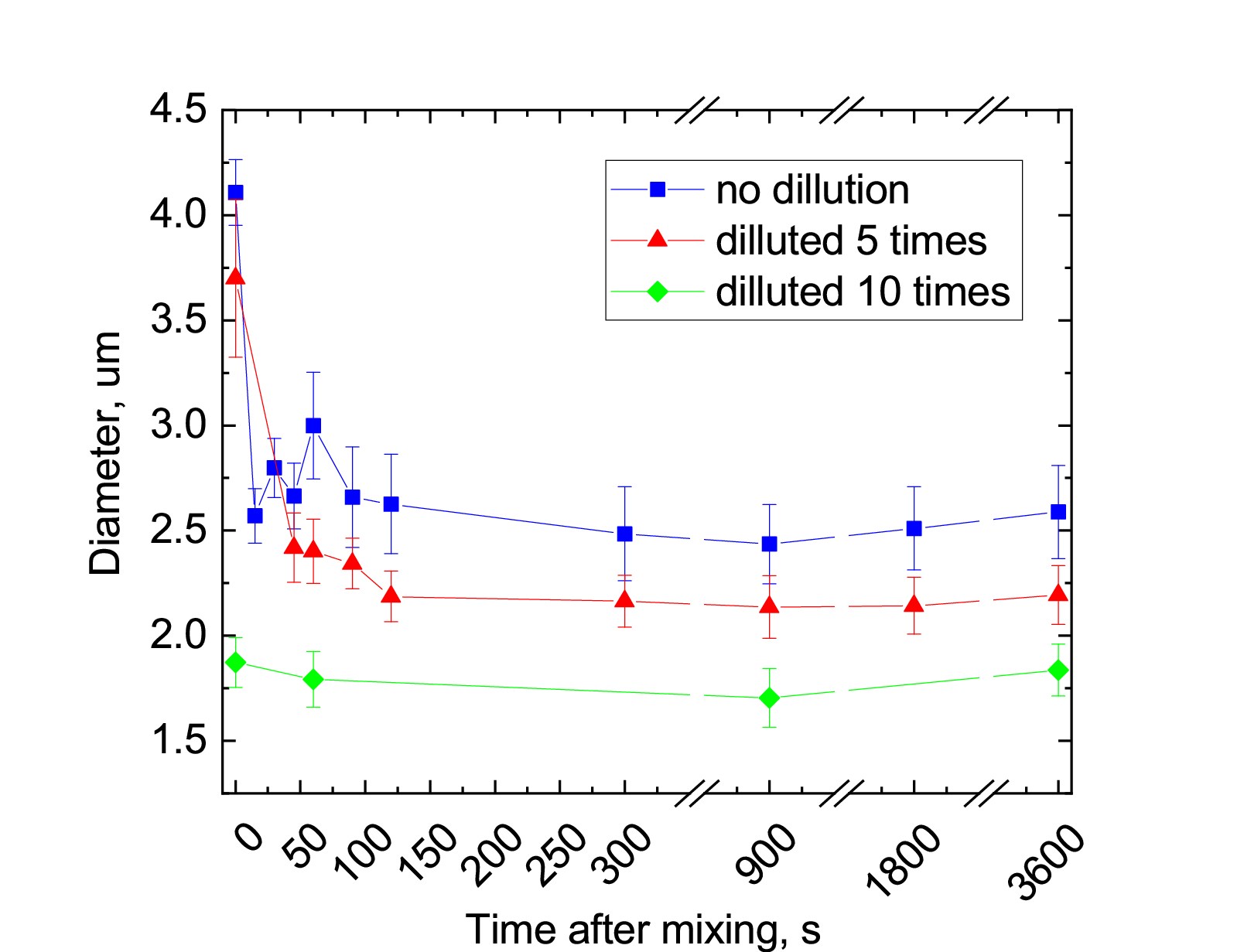}

\caption{Dependencies of calcium carbonate microsphere sizes on the time after mixing and the degree of precursor mixture dilution (undiluted - curve with blue squares, diluted by factors of 5, and 10 - curve with red triangles, green diamonds, respectively).}
\label{fig:3}
\end{figure}

As observed from the curves presented in Figure 3 dilation of reagent mixture in 2, 4 and 10 times effectively decrease the size of resultant \ce{CaCO3} particles up to 3 times of original size (1.7 \(\mu\)m for 10 times diluted solution against 4 um particles of undiluted reagent mixture). A consistent trend of gradual particle size reduction over time after mixing is evident across nearly all samples and can be attributed to redistribution of free ions within the precursor mixture, inducing calcium carbonate crystal rearrangement \cite{huang2022hard}.  Based on the obtained data, the dilution of reagent mix after precursor mixing and prior precipitation from residual reagent molecules emerges as a significant synthesis parameter. Additionally, data presented on Figure 3 shows that implementation of resting time between dilation and centrifugation further decrease particles diameter. For precursors diluted by less than an order of magnitude, a resting time of approximately 100 – 120 seconds results in the reduction of the particle size by 20 – 30\% (from \(\sim \)4 \(\mu\)m to \(\sim \)2.3 \(\mu\)m). 
Based on this data we can expect that precursor mixture transferred from microfluidic chip to external flask for further precipitation does not result in uncontrolled particles size grow. The implementation of an efficient mechanism for partial and/or complete inhibition of calcium carbonate particle synthesis is crucial when transitioning from bulk synthesis to microfluidic chips with high levels of automation.

\subsection{Reagent Mixing Efficiency in microfluidic chip}

In the preceding section, we explored the impact of precursor mixture dilution on the diameter of precipitated particles. The established interaction provided the basis for formulating a synthesis protocol for calcium carbonate particles within microfluidic chips. In this protocol, the final particle size is predominantly influenced by the geometric parameters of the chip and the fluid flow velocities within it. To implement microfluidic synthesis, two microfluidic chips were fabricated using MSLA 3D printing, as detailed in section 2.3 of the Materials and Methods. Both chips maintained an inner volume of 100 \(\mu\)L, with one chip having a cross-section of 0.5 mm\(^2\) and the other 2 mm\(^2\) (Schematic of used microfluidic chips presented in Figure 4a). 
Initially, we investigated the impact of directly transferring the devised synthesis protocol from a flask to a microfluidic chip on the physical properties of the particles. For this analysis, reagents were introduced into the microfluidic chip at a flow speed of 0.001 mL/s, ensuring that the reaction time of the reagent mixture matched the 100 seconds employed in the flask synthesis discussed in the previous section. Figure 4 illustrates confocal images of the synthesized particles along with their size distributions.

\begin{figure}[h!]

\centering

\begin{minipage}{.5\textwidth}
  \centering

 \hspace{-1.2cm}   (\textbf{a})

 \hspace{-1.2cm}   \includegraphics[width=0.8\linewidth]{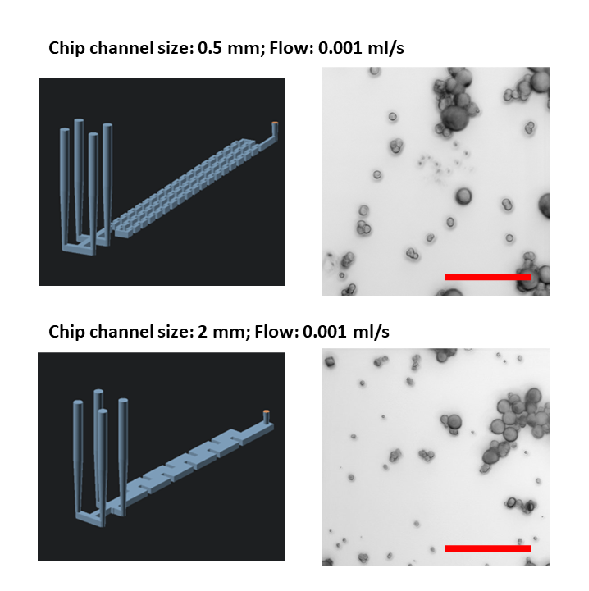}

  \label{fig:4a}
\end{minipage}%
\begin{minipage}{.5\textwidth}
  \centering

    (\textbf{b})

  \includegraphics[width=1.0\linewidth]{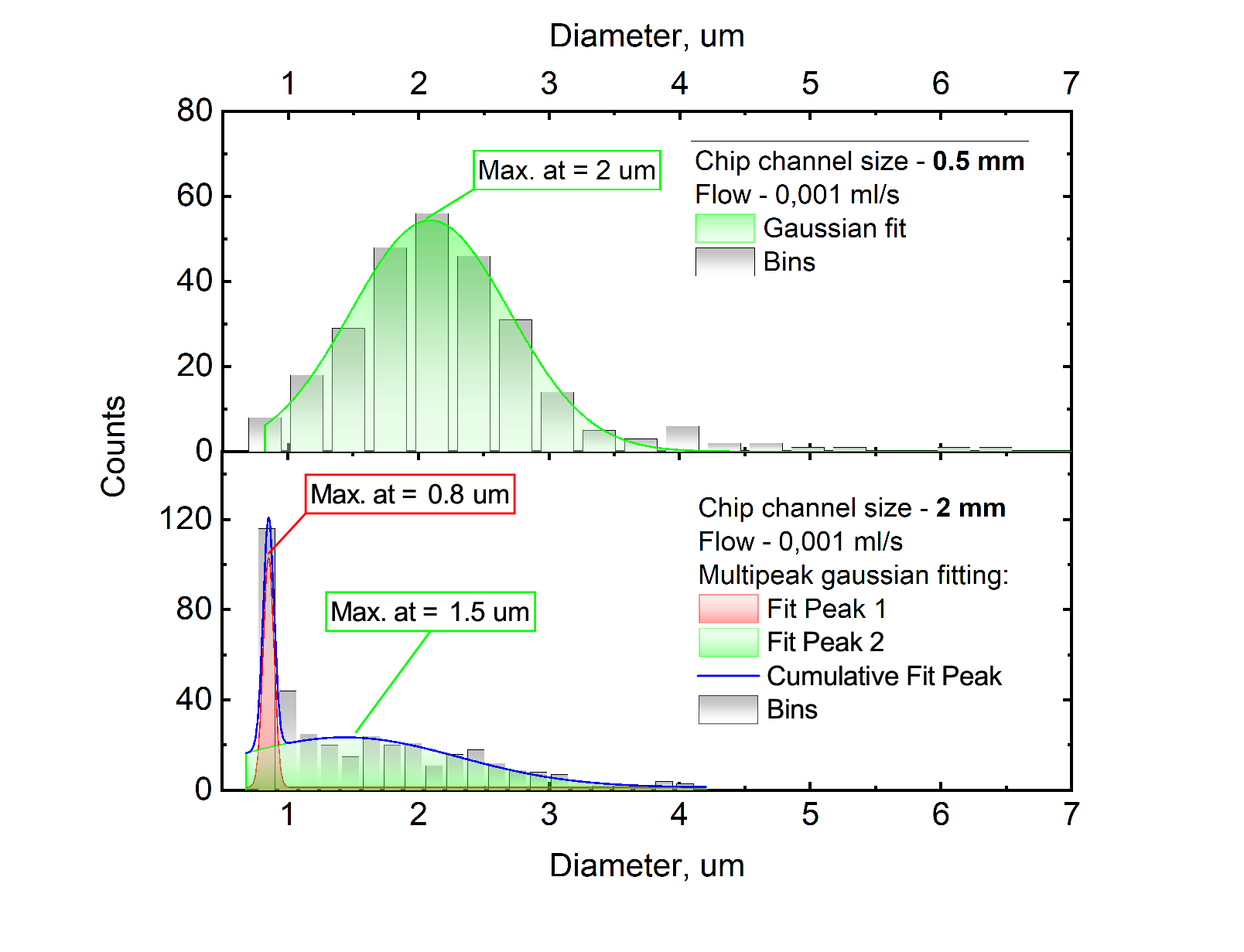}
  
  \label{fig:4b}
\end{minipage}
\caption{(\textbf{a}) $-$ Images of synthesized with different flow speed and chip’s channel width \ce{CaCO3} particles taken by confocal microscopy. (\textbf{b}) $-$ Size distribution of synthesized \ce{CaCO3} particles synthesized with different chip’s width (0.5 and 2 mm for upper and bottom panels, respectively) and reagent flow speed of 0.001 ml/s.}
\label{fig:4}
\end{figure}

The analysis of \ce{CaCO3} particle size distributions, as presented in Figure 4b, reveals the synthesis of a polydisperse ensemble using both microfluidic chip designs. Notably, the size distribution significantly varies between the two chip designs. The 0.5 mm\(^2\) cross-section chip exhibits a broad distribution of micron-sized particles around 2 \(\mu\)m, while the 2 mm\(^2\) cross-section chip displays two distinct peaks at 0.8 and 1.5 \(\mu\)m (Figure 4b, bottom panel). It is noteworthy that submicron particles demonstrate greater uniformity compared to their micron-sized counterparts.

This observed polydispersity and the presence of distinct ensembles in microfluidically synthesized \ce{CaCO3} particles, both submicron and micron-sized, may be attributed to two distinct regimes of reagent mixing within microfluidic channels. Initially, turbulent mixing occurs as the reagent solution traverses stagnant media within the microfluidic channels. This turbulent flow regime arises from the interaction of reagent molecules, moving at high flow speeds, with immobile water molecules pre-existing in the microfluidic chips before \ce{CaCO3} particle synthesis. Following a brief duration, the turbulent regime transitions to a laminar flow of reagent molecules, propelling all molecules within the media towards the outlet. In the laminar regime, a thin area near the center of the microfluidic channel exhibits significant variations in synthesis parameters, especially concentrations of \ce{CaCl2} and \ce{Na2CO3} salts, and their stoichiometric ratio. These factors have been previously shown to significantly influence the growth of \ce{CaCO3} crystals \cite{reznik2023microfluidic}.

During the turbulent regime, uncontrollable mixing of the reagent solution across the microfluidic channel results in the formation of relatively small micron-sized particles with a broad size distribution. In contrast, the laminar regime promotes the formation of submicron particles due to an elevated rate of nucleation. As the rate of nucleation increases and the critical nucleus size decreases, stable nucleus formation prevails over subsequent growth, following Ostwald’s rule of selection. This shift is attributed to an escalation in supersaturation, leading to a higher rate of viable nucleus formation. To support this hypothesis, computational fluid dynamics (CFD) simulations of reagent mixing were conducted for two distinct microfluidic chip configurations, as depicted in Figure 5, illustrating reagent concentration during the synthesis of \ce{CaCO3} particles in microfluidic chips with cross-sectional channel areas of 0.5 and 2 mm\(^2\).

\begin{figure}[h!]

\centering

\begin{minipage}{.5\textwidth}
  \centering

 \hspace{-1.2cm}   (\textbf{a})

 \hspace{-1.2cm}   \includegraphics[width=1.0\linewidth]{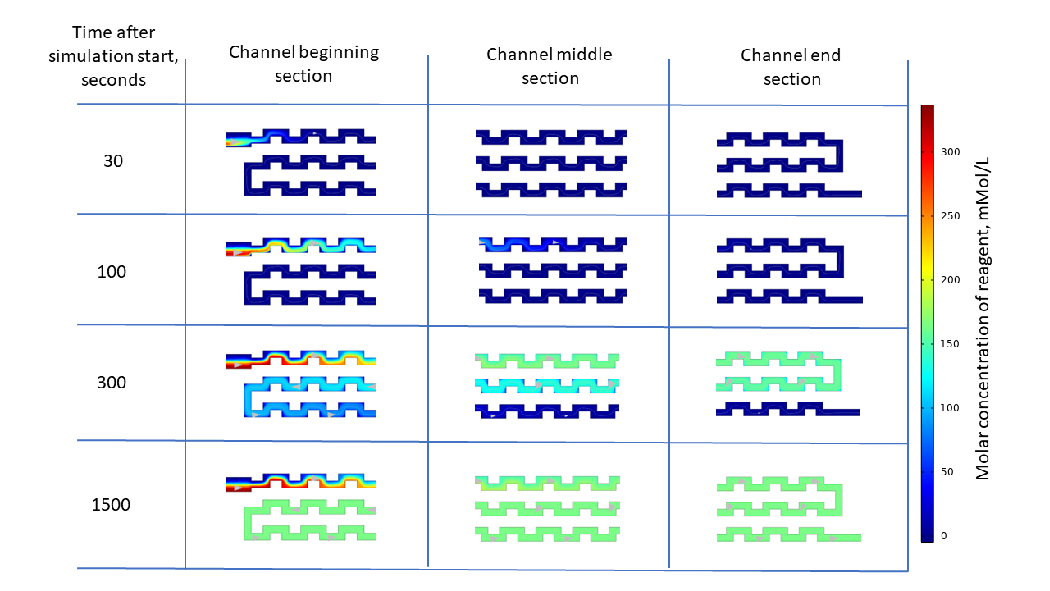}

  \label{fig:5a}
\end{minipage}%
\begin{minipage}{.5\textwidth}
  \centering

    (\textbf{b})

  \includegraphics[width=1.0\linewidth]{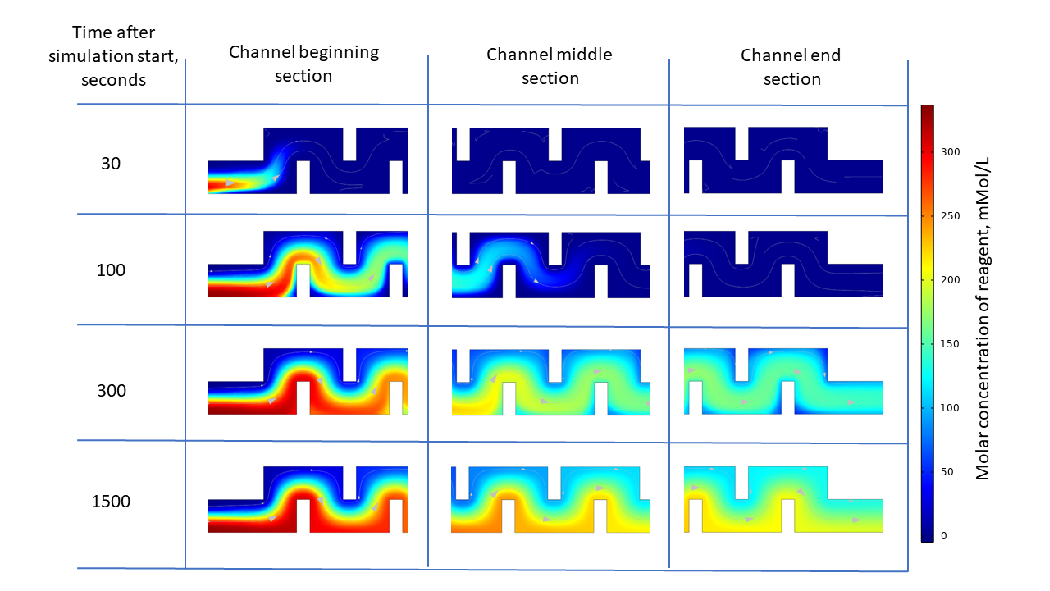}
  
  \label{fig:5b}
\end{minipage}
\caption{CFD simulation of reagent mixing in different microfluidic chip sections, for channel cross section of 0.5 mm\(^2\) (\textbf{a}) and 2 mm\(^2\) (\textbf{b}) at different time frames after start of pumping cycle.}
\label{fig:5}
\end{figure}

In Figure 5a, the reagent distribution within the microfluidic chip appears predominantly homogeneous after 300 seconds of continuous reagent pumping. However, a concentration gradient near the main inlet is noticeable at the initiation of reagent pumping, persisting throughout the simulation period. It is crucial to emphasize that this gradient remains stable over time, in contrast to the simulation depicted in Figure 6b, representing a chip with a wider channel. In the case of wider channels, a gradual transition from turbulent flow during the initial phase of the simulation is observed, followed by the progressive establishment of laminar flow after 300 seconds of continuous pumping. This behavior aligns with the previously posited assumption where a chip with a smaller cross-section primarily generates turbulent mixing, leading to the synthesis of micron-sized particles with a broad size distribution (Figure 4b, top panel). Conversely, a chip with an increased cross-section produces two distinct size ensembles of submicron and micron-sized particles, generated along the center of the channel where a high concentration of both reagents is present and near the channel walls, respectively.

To empirically validate the occurrence of two distinct regimes of reagent flow within a microfluidic chip resulting in such behavior, a series of syntheses were conducted in both chips. In these syntheses, the volume of the reagent solution exiting the microfluidic chip was incrementally increased, starting from 50 \(\mu\)L up to 1000 \(\mu\)L. After traversing the internal volume of the chip, the sampled precursor mixture, along with the formed particles, underwent a tenfold dilution with distilled water and was subsequently subjected to two washes to eliminate unreacted reagent molecules. The parameters for the synthesis of calcium carbonate microspheres in this experiment were the same as the ones described in Figure 4.

Figure 6 presents particle size distribution plots of calcium carbonate microspheres synthesized in microfluidic chips with channel cross-sections of 0.5 and 2 mm\(^2\) and varying amounts of exiting reagent mix.

\begin{figure}[h!]
	
	\centering
	
	\begin{minipage}{.5\textwidth}
		\centering
		
		\hspace{-1.2cm}   (\textbf{a})
		
		\hspace{-1.2cm}   \includegraphics[width=0.9\linewidth]{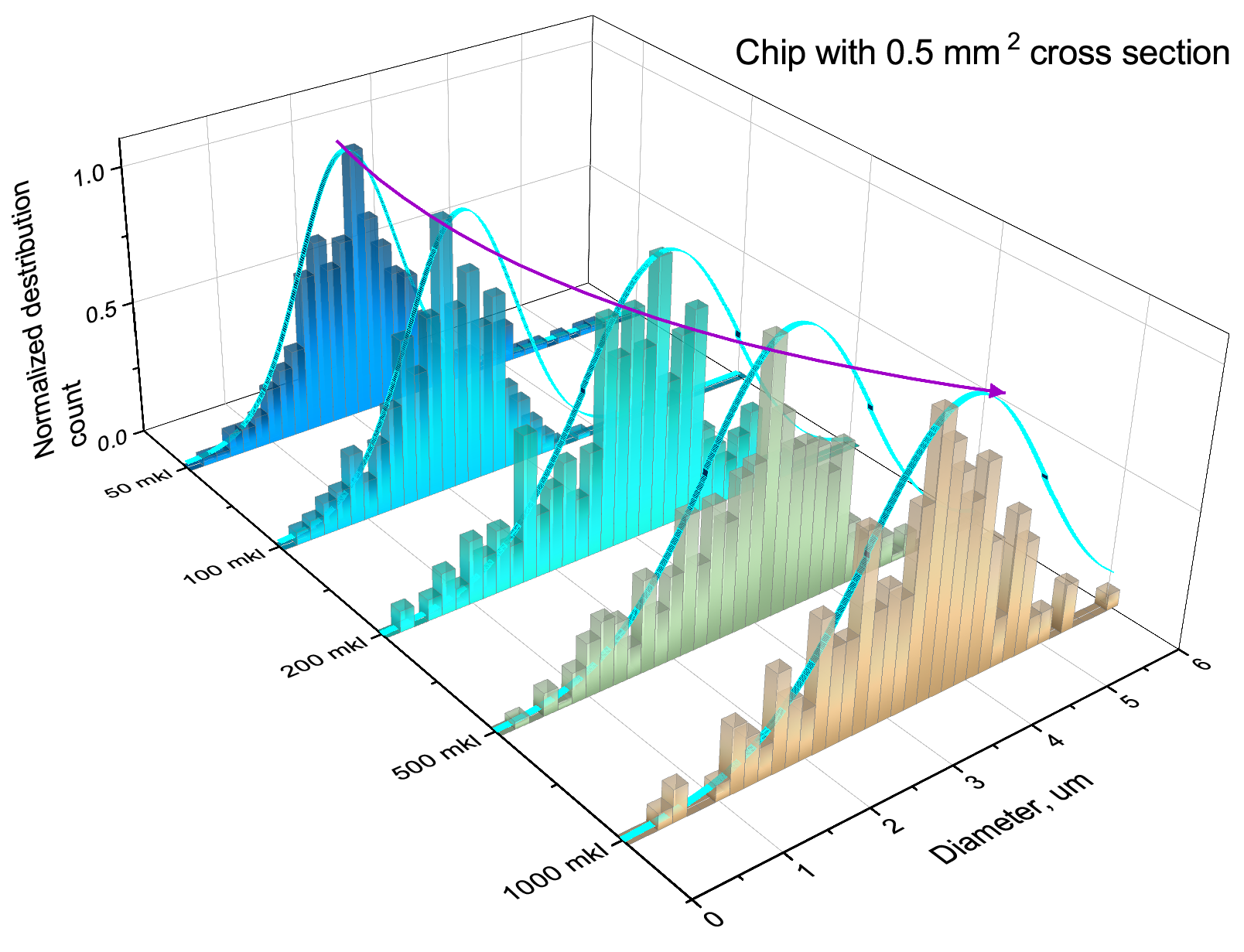}
		
		\label{fig:6a}
	\end{minipage}%
	\begin{minipage}{.5\textwidth}
		\centering
		
		(\textbf{b})
		
		\includegraphics[width=0.9\linewidth]{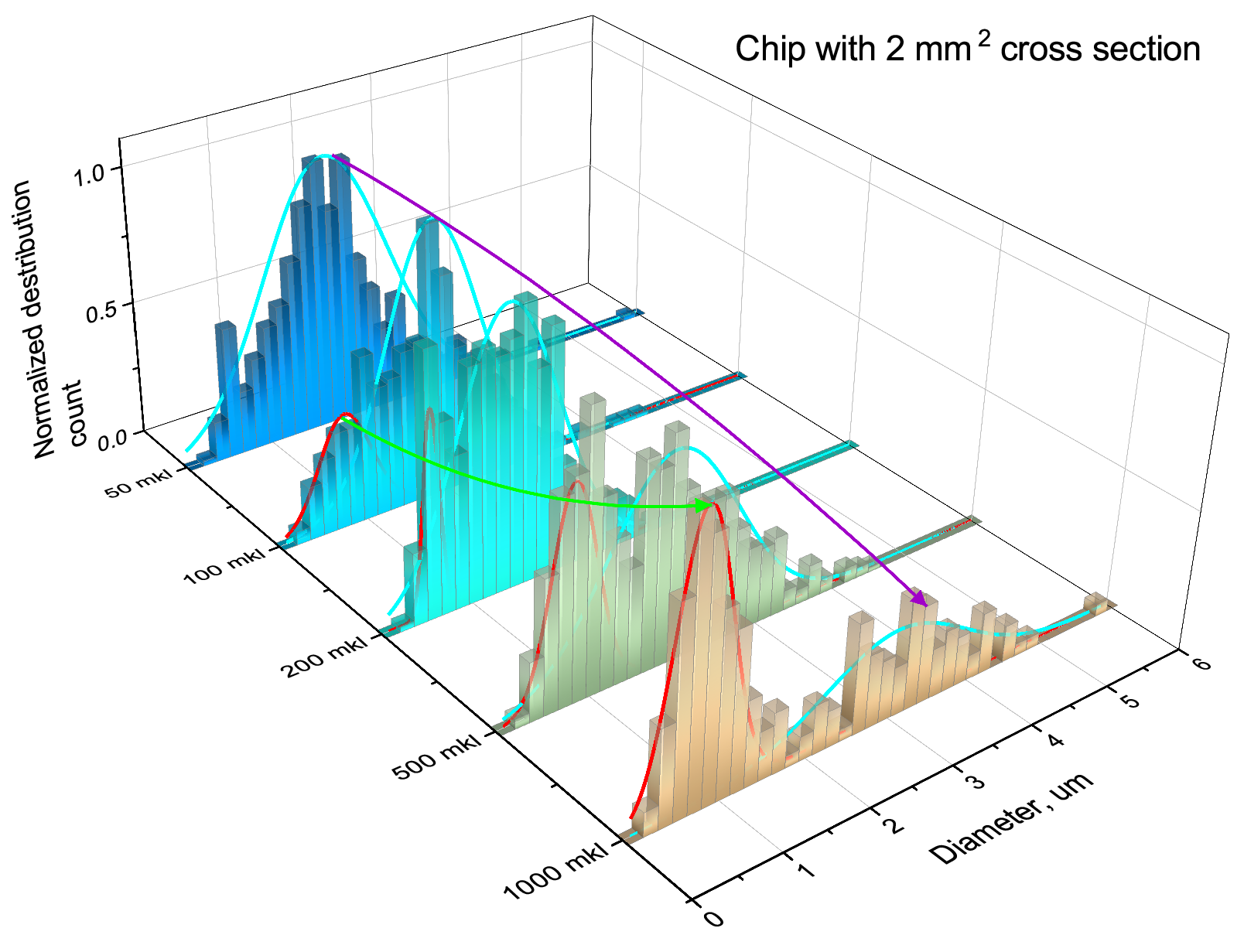}
		
		\label{fig:6b}
	\end{minipage}
	\caption{\ce{CaCO3} size distribution after synthesizing in the microfluidic chip with channel cross section of 0.5 mm\(^2\) (\textbf{a}) and 2 mm\(^2\) (\textbf{b}) in dependence of amount of reagent mix volume that comes out of the chip.}
	\label{fig:6}
\end{figure}

As illustrated in Figure 6, a gradual increase in the volume of the reagent mixture exiting the outlet of the microfluidic chip results in a discernible alteration in the size distribution of the synthesized \ce{CaCO3} particles. Specifically, a slow emergence of a second, narrower peak in the submicron range is observed, with a maximum at 0.9 \(\mu\)m, in addition to the original broader peak. We posit that this progression may be attributed to the previously discussed presence of two reagent mixing regimes, gradually transitioning from turbulent to laminar, influencing the polydispersity of \ce{CaCO3} particles.

Throughout the synthesis of \ce{CaCO3} particles in microfluidic chips, a submicron ensemble is formed. According to the size distribution dependencies presented in Figure 4b and Figure 6d, submicron \ce{CaCO3} particles constitute a substantial portion, ranging from 50\% to 80\% of the total number of particles. Notably, a considerable quantity of submicron particles is produced using the microfluidic synthesis method, approaching the diffractive resolution limit of the confocal microscope (\(\sim \)200 nm). To facilitate a more comprehensive analysis, samples of submicron particles were subjected to investigation using scanning electron microscopy (SEM) and Energy-dispersive X-ray spectroscopy (EDX). Figure 7 presents the SEM and EDX images obtained for the nanoscale \ce{CaCO3} particles synthesized within a microfluidic chip.

\begin{figure}[h!]
	
	\centering
	
	\begin{minipage}{.5\textwidth}
		\centering
		
		\hspace{-1.2cm}   (\textbf{a})
		
		\hspace{-1.2cm}   \includegraphics[width=0.8\linewidth]{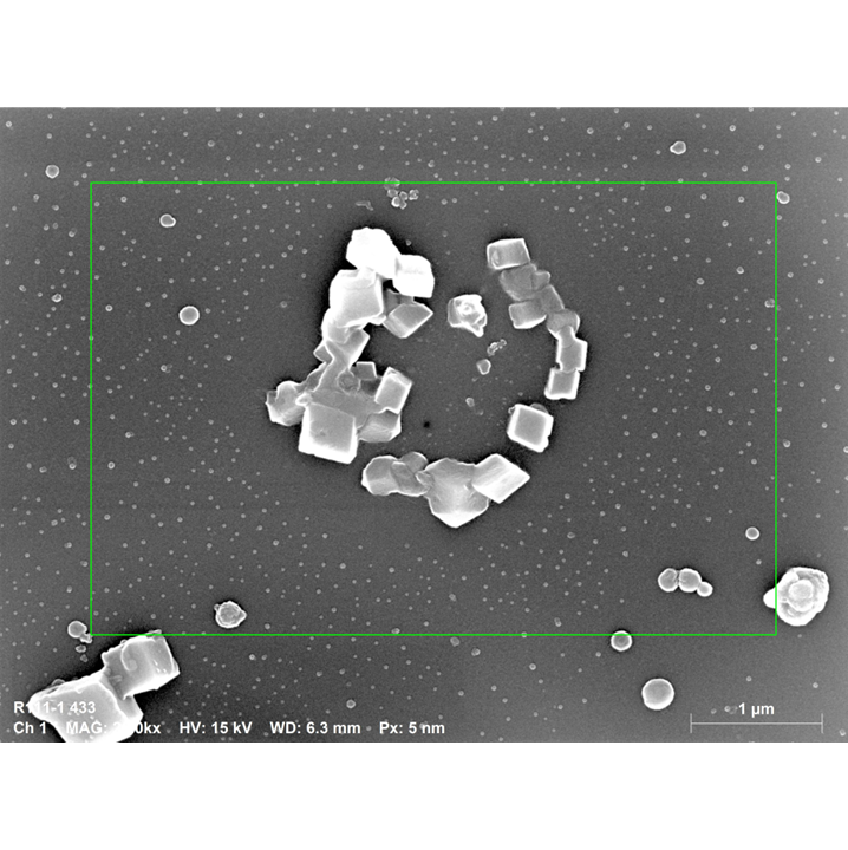}
		
		\label{fig:7a}
	\end{minipage}%
	\begin{minipage}{.5\textwidth}
		\centering
		
		(\textbf{b})
		
		\includegraphics[width=0.9\linewidth]{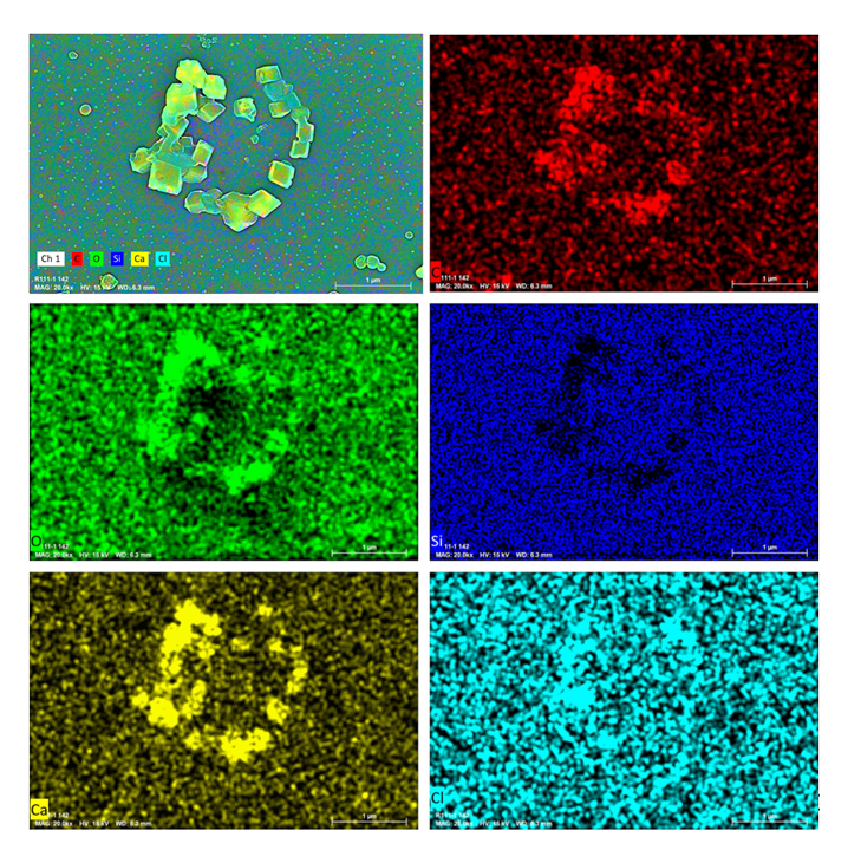}
		
		\label{fig:7b}
	\end{minipage}
	\caption{SEM (\textbf{a}) and EDX (\textbf{b}) images of \ce{CaCO3} submicron particles deposited on to Si wafer. Colors on EDX images represent recorded signal from various elements as follows: Red - for Carbon; Green - Oxygen; Blue - Silicon; Yellow - Calcium; Cyan - Chloride. }
	\label{fig:7}
\end{figure}

As depicted in Figures 7a and 7b, particles synthesized within the microfluidic chip exhibit a considerable presence of nanosized particles ranging around 50-90 nm in diameter. Elemental analysis using Energy-dispersive X-ray spectroscopy (EDX) confirms the composition of these particles, revealing the presence of Calcium, Carbon, and Oxygen atoms, with slight traces of Chloride. Notably, while micron sized \ce{CaCO3} particles predominantly adopt the vaterite form, submicron particles exhibit a more prevalent occurrence of the calcite form. This phenomenon may be influenced by residual \ce{CaCl2} reagent traces (detected on the particles), which were not entirely removed before the deposition of particles on a wafer. The presence of these residual reagent molecules could have induced the recrystallization of particles from the vaterite to the calcite form \cite{beck2010spherulitic}.

The data presented in this section serves to validate that the utilization of 3D printed microfluidic chips with relatively wide channel diameters can yield high-quality submicron and even nanoscale particles. The findings affirm that microfluidic synthesis of calcium carbonate microspheres is not only achievable but also surpasses classical flask synthesis in terms of control. Notably, the microfluidic-produced calcium carbonate particles exhibit a submicron size, a highly desirable characteristic for potential applications in biomedicine. Furthermore, the study demonstrates that microfluidic chips can be directly fabricated using additive manufacturing methods, eliminating the necessity for traditional soft lithography.

\section{Conclusion}

This investigation delves into the intricacies of calcium carbonate synthesis employing a microfluidic approach. The utilization of 3D-printed microfluidic chips in our experiments underscored their appropriateness for achieving precision and control in \ce{CaCO3} particle synthesis. Our endeavor to achieve smaller particle sizes, achieved through the modification of both synthesis and post-synthesis procedures, unveiled a nuanced equilibrium between reducing particle size and maintaining synthesis reproducibility. 

The comprehensive analysis of particle size distribution, coupled with the investigation of varying reagent volumes exiting the microfluidic chip, indicates a noteworthy alteration in the size distribution of the synthesized particles. The emergence of a second, narrower peak in the submicron range suggests a nuanced relationship between reagent flow dynamics and particle size distribution. The observed transition from turbulent to laminar flow regimes, aligns with the proposed hypothesis of two distinct mixing regimes affecting the polydispersity of \ce{CaCO3} particles.

Furthermore, the microscopic examination, along with EDX analysis, reveals the presence of a substantial quantity of nanosized particles (50-90 nm in diameter) within the synthesized particles. Notably, the study highlights the impact of residual reagent traces, particularly \ce{CaCl2}, on the crystalline forms of the synthesized particles. The shift from predominantly vaterite in micron-sized particles to calcite in submicron particles suggests the potential role of residual reagents in triggering recrystallization.

Overall, the findings affirm the efficacy of microfluidic synthesis in producing high-quality submicron and nanoscale particles, demonstrating superior control compared to classical flask synthesis. The submicron size of the microfluidic-produced calcium carbonate particles holds promise for applications in biomedicine. Moreover, the utilization of 3D printed microfluidic chips with wide channel diameters presents an innovative approach, eliminating the need for traditional soft lithography in chip fabrication. This research contributes valuable insights to the understanding and optimization of microfluidic synthesis processes for the controlled production of submicron and nanoscale particles.

\section*{Author Contributions}
Reznik I.A.: conceptualization, methodology, investigation, validation, visualization, writing – original draft; Kolesova E.P.: writing – review and editing; Pestereva A.S.: methodology, investigation; Baranov K.N.: methodology, investigation; Osin Y.N.: methodology, investigation; Bogdanov K.V.: methodology, investigation; Swart J.W.: writing – review and editing, Moshkalev S.A.: writing – review and editing; Orlova A.O.:
funding acquisition, validation, supervision.

\section*{Conflicts of interest}
The authors declare no competing financial interest. 

\section*{Acknowledgments}
This work was financially supported by the Ministry of Education and Science of the Russian Federation, State assignment, Passport 2019-1080 (Goszadanie 2019-1080).

\bibliographystyle{unsrt}  
\bibliography{CaCO3_Reznik}

\end{document}